\documentclass[aps,prd,twocolumn,a4paper,superscriptaddress,floatfix]{revtex4-1}

\usepackage{graphicx}
\usepackage{amssymb}
\usepackage{natbib}
\usepackage{color}
\usepackage[dvipsnames]{xcolor}
\usepackage{amsmath}    
\usepackage{mathrsfs}   
\usepackage{hyperref}

\DeclareFontFamily{U}{mathx}{\hyphenchar\font45}
\DeclareFontShape{U}{mathx}{m}{n}{<-> mathx10}{}
\DeclareSymbolFont{mathx}{U}{mathx}{m}{n}

\DeclareMathAccent{\widebar}{0}{mathx}{"73} 

\begin{document}

\newcommand\lsim{\mathrel{\rlap{\lower4pt\hbox{\hskip1pt$\sim$}}
 	\raise1pt\hbox{$<$}}}
\newcommand\gsim{\mathrel{\rlap{\lower4pt\hbox{\hskip1pt$\sim$}}
 	\raise1pt\hbox{$>$}}}

\def\app#1#2{%
  \mathrel{%
    \setbox0=\hbox{$#1\sim$}%
    \setbox2=\hbox{%
      \rlap{\hbox{$#1\propto$}}%
      \lower1.1\ht0\box0%
    }%
    \raise0.25\ht2\box2%
  }%
}
\def\approxprop{\mathpalette\app\relax}


\title{Lagrangian Identity and Mass Evolution of Particle-like Objects in Nonminimally Coupled Gravity}

\author{S. R. Pinto}
\email{samuel.pinto@astro.up.pt}
\affiliation{Departamento de Física e Astronomia, Faculdade de Ci\^encias, Universidade do Porto, Rua do Campo Alegre, 4169-007 Porto, Portugal}
\affiliation{Instituto de Astrof\'{\i}sica e Ci\^encias do Espa\c co, CAUP, Rua das Estrelas, 4150-762 Porto, Portugal}
\author{P. P. Avelino}
\email{pedro.avelino@astro.up.pt}
\affiliation{Departamento de Física e Astronomia, Faculdade de Ci\^encias, Universidade do Porto, Rua do Campo Alegre, 4169-007 Porto, Portugal}
\affiliation{Instituto de Astrof\'{\i}sica e Ci\^encias do Espa\c co, CAUP, Rua das Estrelas, 4150-762 Porto, Portugal}

\begin{abstract}

We show that the Lagrangian of a Nambu-Goto $p$-brane satisfies the identity $\mathcal{L}_{\rm [\it p \rm]}=T_{\rm [\it p \rm]}/(p+1)$, with $T_{\rm [\it p \rm]}$ denoting the trace of the corresponding energy-momentum tensor, independently of the properties of the gravitational field. While for $p=0$ this reduces to the standard $\mathcal{L}_{\rm [0]}=T_{\rm [0]}$ relation, which determines the on-shell Lagrangian of point particles and their fluids, more generally it depends explicitly on the $p$-brane dimensionality. We explore the implications of this Lagrangian identity for the dynamics of non-self-intersecting cosmic string loops in a homogeneous and isotropic universe within nonminimally coupled scalar-tensor gravity, showing that, unlike in general relativity, their rest mass can evolve in response to the cosmological evolution of the background spacetime, regardless of their small size or tension. We further generalize this analysis to closed $p$-branes in $(N+1)$-dimensional Friedmann-Lemaître-Robertson-Walker spacetimes, showing that the evolution of the rest mass depends explicitly on the dimensionality of the brane, and therefore that the cosmological evolution of particle-like objects in theories of gravity with nonminimal matter couplings is sensitive to their internal structure.

\end{abstract}
\date{\today}
\maketitle

\section{Introduction \label{sec1}}

In General Relativity (GR), the on-shell matter Lagrangian does not enter explicitly into the gravitational field equations. Therefore, distinct on-shell matter Lagrangians that correspond to the same energy-momentum tensor are physically equivalent within GR. This degeneracy reflects the fact that distinct on-shell matter Lagrangians contribute identically to the Einstein equations. The situation changes in theories featuring a nonminimal coupling between matter and gravity \cite{Harko:2010mv, Harko:2011kv, Bertolami_2007, Capozziello:2011et, Clifton:2011jh, Haghani:2013oma, Katirci:2013okf, Ludwig_2015, Nojiri:2017ncd, Bahamonde:2017ifa, Avelino:2018rsb, Harko:2018gxr, Azevedo:2018nvi, Minazzoli:2018xjy, Azevedo:2019krx, Fisher:2019ekh, Avelino:2020fek, Arruga:2020knb, Azevedo:2021npm, Arruga:2021qhc, Carvalho:2022kxq, Pappas:2022gtt, Goncalves:2023umv, Jana:2023djt, Lacombe:2023pmx, Harko:2024sea, Asghari:2024obf, Asghari:2024qgp, Kaczmarek:2024quk, Errehymy:2025kzj, Minazzoli:2025gyw, Boehmer:2025afy, Olmo:2025wot,Chehab:2026obk}. In these theories, the on-shell matter Lagrangian can appear explicitly in the gravitational action and contribute directly to the field equations. Consequently, different on-shell matter Lagrangians, even if they correspond to the same energy-momentum tensor in GR, generally lead to inequivalent dynamics. The degeneracy present in GR is therefore lifted, and the precise form of the on-shell matter Lagrangian acquires physical significance.

Perfect fluids provide a widely used and sufficiently general description of the sources of the gravitational field in many astrophysical and cosmological contexts. The Lagrangian formulation commonly employed to describe perfect fluids \cite{Schutz:1970my,Ray:1972,Schutz:1977df,Taub:1978,Matarrese:1984zw,Brown:1992kc, Andersson:2006nr, Minazzoli:2012md,Ferreira:2020fma} offers a consistent macroscopic framework that reproduces the standard hydrodynamic equations. However, this approach does not explicitly account for the microscopic particle dynamics underlying the fluid description. In fact, within GR, several equivalent Lagrangian formulations exist for a perfect fluid at the macroscopic level, leading to different on-shell Lagrangians \cite{Brown:1992kc,Ferreira:2020fma}. These distinct forms are physically indistinguishable in minimally coupled gravity, as long as they yield the same energy-momentum tensor. However, in the presence of a nonminimal coupling, this equivalence generally breaks down. Since the matter Lagrangian now enters explicitly into the field equations, both gravitational and matter dynamics may depend significantly on the specific on-shell matter Lagrangian. 

This raises the fundamental question of whether a preferred, physically motivated, on-shell Lagrangian can be identified. Indeed, it has been shown that no universal on-shell Lagrangian exists even for a perfect fluid \cite{Avelino:2018rsb,Ferreira:2020fma}. Nevertheless, in the case of an ideal gas, the on-shell matter Lagrangian has been shown to be equal to the trace of the energy-momentum tensor \cite{Avelino:2018rsb,Ferreira:2020fma,Avelino:2018qgt,Avelino:2022eqm,Pinto:2025plg}. This result relies crucially on microscopic considerations, in particular the  assumptions that the constituent particles possess fixed rest mass and internal structure, and that their self-gravity can be neglected. However, in systems endowed with internal degrees of freedom, such as the oscillatory modes of a $1$-brane, the assumption of fixed structure no longer holds.

In this work, we investigate how relaxing this assumption in the presence of a nonminimal coupling affects the evolution of the rest mass of particle-like objects, namely closed, non-self-intersecting cosmic string loops and, more generally, $p$-branes. The structure of the paper is as follows. In Sec.~\ref{ST_theories}, we present the field equations for the gravitational and matter fields in the context of a broad class of scalar-tensor theories nonminimally coupled to matter. We also discuss two important subclasses, corresponding to nonminimally coupled Brans-Dicke theory and $f(R, \mathcal L_{\rm m})$ gravity. In Sec.~\ref{branes}, we derive a general Lagrangian identity relating the Lagrangian of a $p$-brane to the trace of its energy-momentum tensor, thereby generalizing the standard result for point particles ($p=0$). In Sec.~\ref{sec3}, we consider periodic, non-self-intersecting oscillating string loops (closed 1-branes) in Minkowski spacetime and derive a relation between their rest mass and the spatial volume integral of their time-averaged Lagrangian. In Sec.~\ref{cosmological}, we investigate the cosmological evolution of the rest mass of small cosmic string loops in nonminimally coupled gravity, assuming that the string tension is sufficiently small, so that gravitational radiation losses can be neglected. In Sec.~\ref{cosmological2}, we extend these results to closed $p$-branes evolving in an $(N+1)$-dimensional FLRW background. Finally, in Sec.~\ref{concl}, we summarize our results and present our conclusions.

Throughout this work, we adopt the metric signature $[-, +, +, +]$ and use natural units where $c = 16\pi G =1$. The Einstein summation convention is employed, meaning that when an index variable appears twice in a single term, once in an upper (superscript) and once in a lower (subscript) position, it implies summation over all possible values of the index. Unless stated otherwise, Greek indices take the values $0,\dots,N$, the Latin indices $i$ and $j$ take the values $1,\dots,N$, and the Latin indices $a$, $b$ and $c$ take the values $0,\dots,p$.

\section{Nonminimally coupled Scalar-tensor theories \label{ST_theories}}

In this section, we derive the field equations governing the dynamics of the gravitational and matter fields in a broad class of of nonminimally coupled scalar-tensor theories, defined by the action
\begin{eqnarray}
S&=&\int d^4x\sqrt{-g}\Big[ \phi R+\psi \mathcal{L}_{\rm m}-\frac{\omega}{\phi}\nabla_\mu\phi\nabla^\mu\phi \nonumber \\
&-&K(\psi)\nabla_\mu\psi\nabla^\mu\psi-V(\phi,\psi,\chi)   \Big]\,,
\label{action_ST}
\end{eqnarray}
where $g \equiv\det(g_{\alpha\beta})$ is the determinant of the metric, $R \equiv g^{\alpha\beta} R_{\alpha\beta}$ is the Ricci scalar constructed from the Ricci tensor $R_{\alpha\beta}$, $\mathcal{L}_{\rm m}$ is the matter Lagrangian, $\phi$, $\psi$ are dynamical scalar fields that are coupled through a potential $V(\phi,\psi,\chi)$, $\chi$ being an auxiliary scalar field, $\omega$ is a dimensionless constant, and $K(\psi)$ is a function of $\psi$. 

Variation of the action with respect to the metric and the scalar fields $\phi$ and $\psi$ yields
\begin{eqnarray}
&&\left[G_{\alpha\beta} - \Delta_{\alpha\beta}\right] \phi-\frac{\omega}{\phi}\nabla_\alpha\phi\nabla_\beta\phi -K\nabla_\alpha\psi\nabla_\beta\psi \label{Equation_motion_g} \\
&=&\frac{1}{2}\left[\ T_{\alpha\beta}\psi -g_{\alpha\beta}\left(V+\frac{\omega}{\phi}(\nabla\phi)^2+K(\nabla\psi)^2\right)\right] \nonumber\,, \\
&&\nonumber \\
&&R-\frac{\omega}{\phi^2}(\nabla\phi)^2+\frac{2 \omega}{\phi}\Box\phi=V_{,\phi} \label{Equation_motion_phi}\,,\\
&&\nonumber \\
&&\mathcal{L}_{\rm m}+K_{,\psi}(\nabla\psi)^2+2K\Box\psi=V_{,\psi} \label{Equation_motion_psi}\,,
\end{eqnarray}
where $G_{\alpha\beta}\equiv R_{\alpha\beta}-\frac{1}{2}g_{\alpha\beta}R$, $(\nabla\phi)^2\equiv\nabla_\mu\phi\nabla^\mu\phi$, $(\nabla\psi)^2\equiv\nabla_\mu\psi\nabla^\mu\psi$, $\Delta_{\alpha\beta} \equiv \nabla_{\alpha} \nabla_{\beta} - g_{\alpha\beta} \Box$, $\Box \equiv \nabla_{\alpha} \nabla^{\alpha}$, $X_{,\phi} \equiv  \partial X/\partial \phi$, $X_{,\psi} \equiv  \partial X /\partial \psi$ (with $X$ being an arbitrary function of $\phi$, $\psi$ and $\chi$), and $T_{\alpha\beta}$ are the components of the energy-momentum tensor defined by 
\begin{equation}
T_{\alpha\beta} \equiv -\frac{2}{\sqrt{-g}} \frac{\delta (\sqrt{-g} \mathcal{L}_{\rm m})}{\delta g^{\alpha\beta}} \,.
\label{Definition_EMT}
\end{equation}

Unlike in GR, where $\nabla_\beta {T_\alpha}^{\beta}=0$ follows directly from the contracted Bianchi identities, the action defined in Eq.~\eqref{action_ST} contains a nonminimal coupling between $\psi$ and $\mathcal{L}_{\rm m}$, which generally leads to a violation of the local conservation of energy and momentum. Combining Eqs.~\eqref{Equation_motion_g}--\eqref{Equation_motion_psi}, we obtain
\begin{equation}
\nabla_\beta {T_{\alpha}}^\beta=\frac{\nabla_\beta \psi}{\psi}\left(\mathcal{L}_{\rm m}{\delta_\alpha}^\beta-{T_{\alpha}}^\beta\right)\,,
\label{Continuity_Eq}
\end{equation}
where ${\delta_\alpha}^\beta$ denotes the Kronecker delta. Equation~\eqref{Continuity_Eq} shows that, unlike in GR, the matter dynamics explicitly depend on the on-shell matter Lagrangian.

\subsection{Nonminimally coupled Brans-Dicke gravity}

The theory described by the action presented in Eq.~\eqref{action_ST} encompasses, as particular cases, several well-behaved scalar-tensor theories~\cite{Brans, Minazzoli_2014, Brax_2010}. One important example is Brans-Dicke theory with a universal nonminimal coupling between the scalar field and the matter Lagrangian~\cite{Minazzoli_2014}. This theory is obtained by choosing the potential to be of the form
\begin{equation}
    V(\phi,\psi,\chi)=W(\phi)-\chi[\psi-h(\phi)]\,,
    \label{Portential_BD}
\end{equation}
where the auxiliary Lagrange multiplier $\chi$ enforces the constraint
$\psi=h(\phi)$, with $h(\phi)$ being a dimensionless function of $\phi$. Substituting this potential into Eq.~\eqref{action_ST}, together with
$K(\psi)=0$, and imposing the constraint $\psi=h(\phi)$, the action becomes
\begin{eqnarray}
    S=\int d^4x\sqrt{-g}\Big[ \phi R+h(\phi)\mathcal{L}_{\rm m}-\frac{\omega}{\phi}(\nabla\phi)^2-W(\phi)\Big]\,,
    \label{action_BD}
\end{eqnarray}
The equations of motion are given by:
\begin{eqnarray}
&&\left[G_{\alpha\beta} - \Delta_{\alpha\beta}\right] \phi-\frac{\omega}{\phi}\nabla_\alpha\phi\nabla_\beta\phi \label{Equation_motion_g_BD} \\
&=&\frac{1}{2}\left[ T_{\alpha\beta}h(\phi) -g_{\alpha\beta}\left(W+\frac{\omega}{\phi}(\nabla\phi)^2\right)\right] \nonumber\,, \\
&&\nonumber \\
W_{,\phi}&=&R-\frac{\omega}{\phi^2}(\nabla\phi)^2+\frac{2 \omega}{\phi}\Box\phi+h_{,\phi}(\phi)\mathcal{L}_{\rm m} \label{Equation_motion_phi_BD}\,.
\end{eqnarray}
For this subclass of theories, Eq.~\eqref{Continuity_Eq} remains valid upon replacing $\psi$ with $h(\phi)$. 

Models within this subclass are generally well behaved in the sense that they propagate  only the standard tensor and scalar degrees of freedom and are free from Ostrogradsky ghost instabilities. Indeed, the action~\eqref{action_BD} contains no higher-order derivatives, and an invertible conformal transformation maps it to the Einstein frame, where the theory takes the form of General Relativity coupled to a canonical scalar field with a universal matter coupling~\cite{FujiiMaeda2003,Faraoni2004,Minazzoli_2012}. 

It is worth noting that scalar-tensor theories with universal matter couplings naturally arise in the low-energy effective action of string theory, where the scalar field plays the role of a nonminimally coupled dilaton with a self-interaction potential~\cite{Damour_1994, Brax_2010}. Their emergence in this context further supports the theoretical consistency and physical relevance of this class of models.

\subsection{$f(R,\mathcal{L}_{\rm m})$ gravity}

Another important subclass of the action~\eqref{action_ST} corresponds to the scalar-tensor representation of $f(R,\mathcal{L}_{\rm m})$ gravity~\cite{Harko:2010mv}. In this case, the scalar fields $\phi$ and $\psi$ are auxiliary, \emph{i.e.}, they have no kinetic terms ($\omega=0$ and $K=0$), while the potential is independent of $\chi$, so that $V(\phi,\psi,\chi)=W(\phi,\psi)$. The action then reduces to
\begin{equation}
    S=\int d^4x\sqrt{-g}\left[
    R\phi
    +\mathcal{L}_{\rm m}\psi
    -W(\phi,\psi)
    \right]\,.
    \label{action_f(R,L_m)}
\end{equation}
Variation of the action with respect to the scalar fields $\phi$ and $\psi$ yields the conditions
\begin{equation}
    R=W_{,\phi}\,, \qquad
    \mathcal{L}_{\rm m}=W_{,\psi}\,. \label{R,L}
\end{equation}
If the Hessian matrix of $W(\phi,\psi)$ is nonsingular,
\begin{equation}
W_{,\phi\phi}W_{,\psi\psi}-{W_{,\phi\psi}}^2\neq0\,,
\end{equation}
these algebraic equations can be inverted to express the 
scalar fields $\phi$ and $\psi$ as functions of $R$ and $\mathcal{L}_{\rm m}$. Substituting these expressions into Eq. \eqref{action_f(R,L_m)} yields the dynamically equivalent action
\begin{equation}
    S=\int d^4x\sqrt{-g}\,f(R,\mathcal{L}_{\rm m})\,,
\end{equation}
which no longer depends on the scalar fields $\phi$ and $\psi$. By construction, the function $f(R,\mathcal{L}_{\rm m})$ is the Legendre transform of
\begin{equation}
W(\phi,\psi)=\phi R+\psi\mathcal{L}_{\rm m}-f(R,\mathcal{L}_{\rm m})\,.
\end{equation}
By the properties of the Legendre transform,
\begin{equation}
\phi=f_{,R}\,,\qquad
\psi=f_{,\mathcal{L}_{\rm m}}\,.
\end{equation}

In this work, we consider in particular models that are linear in the matter Lagrangian \cite{Bertolami_2007},
\begin{equation}
f(R,\mathcal{L}_{\rm m})
=
f_1(R)+f_2(R)\,\mathcal{L}_{\rm m},
\end{equation}
where $f_1(R)$ and $f_2(R)$ are arbitrary functions of the Ricci scalar. In this case, Eq.~\eqref{Continuity_Eq} holds with the identification $\psi=f_{,\mathcal{L}_{\rm m}}=f_2(R)$, namely,
\begin{equation}
\nabla_\beta {T_{\alpha}}^\beta=
\frac{\nabla_\beta f_2}{f_2}
\left(
\mathcal{L}_{\rm m}{\delta_\alpha}^\beta
-
{T_{\alpha}}^\beta
\right)\,,
\label{Continuity_Eq_2}
\end{equation}
where $f_2\equiv f_2(R)$.

The scalar-tensor representation derived above is dynamically equivalent to $f(R,\mathcal{L}_{\rm m})$ gravity, provided that the Legendre map is invertible. Nevertheless, the physical viability and dynamical consistency of these theories remain the subject of ongoing investigation, with recent work indicating that consistency requirements may severely restrict the class of viable $f(R,\mathcal{L}_{\rm m})$ models~\cite{Lacombe:2023pmx}.

\section{Lagrangian Identity for Nambu-Goto $p$-Branes \label{branes}}

In this section we derive a general Lagrangian identity for Nambu–Goto $p$-branes, establishing a direct relation between the $p$-brane Lagrangian and the trace of the corresponding energy-momentum tensor.

Consider a $p$-brane described by the Nambu-Goto action
\begin{equation}
S_{\rm [\it p \rm]}=-\mu_{\rm [\it p \rm]} \int d^{p+1} \zeta \sqrt{-\det(h_{ab})} \,,
\label{action}
\end{equation}
where $\mu_{\rm [\it p \rm]}$ denotes the brane tension (rest mass per unit $p$-volume), and $\zeta^a$ ($a=0,1,\ldots,p$) are coordinates on the $(p+1)$-dimensional world volume of the brane. The induced metric on the brane is given by
\begin{equation}
h_{ab} = g_{\alpha\beta}\, x^\alpha_{,a}  x^\beta_{,b}\,,
\label{gamma}
\end{equation}
where $x^\alpha(\zeta)$ specifies the embedding of the brane in the $(N+1)$-dimensional background spacetime (with $N > p$), and a comma denotes partial differentiation with respect to the world-volume coordinate $\zeta^a$, i.e., $x^\alpha_{,a} \equiv \partial x^\alpha / \partial \zeta^a$.

The $p$-brane action can equivalently be expressed as an integral over the $(N+1)$-dimensional background spacetime:
\begin{equation}
S_{\rm [\it p \rm]} = \int d^{N+1}x \, \sqrt{-g}\, \mathcal{L}_{\rm [\it p \rm]},
\label{eq:action_density}
\end{equation}
where the matter Lagrangian takes the form
\begin{equation}
\mathcal{L}_{\rm [\it p \rm]} = -\frac{\mu_{\rm [\it p \rm]}}{\sqrt{-g}} \int d^{p+1}\zeta \, \sqrt{-h} \, \delta^{(N+1)}\!\left[x^\sigma - x^\sigma(\zeta)\right],
\label{lagrangian}
\end{equation}
with $h \equiv \det(h_{ab})$ and $\delta^{(N+1)}$ representing the Dirac delta distribution in $(N+1)$ dimensions.

Using Eq.~\eqref{gamma}, the energy-momentum tensor of the $p$-brane is obtained by varying the action with respect to the spacetime metric,
\begin{eqnarray}
T^{\alpha \beta}_{\rm [\it p \rm]}&=& \frac{2}{\sqrt {-g}}  \frac{\delta(\sqrt{-g}\mathcal L_{\rm [\it p \rm]})}{\delta g_{\alpha \beta}} \label{EMtensor} \\
&=& -\frac{\mu_{\rm [\it p \rm]}}{\sqrt {-g}} \nonumber \\
&\times& \int  \sqrt{-h}  h^{ab} x^\alpha_{,a} x^\beta_{,b}    \delta^{(N+1)} \left[x^\sigma - x^\sigma (\zeta)\right]d^{p+1} \zeta \nonumber\,,
\end{eqnarray}
where $h^{ab}$ denotes the inverse induced metric satisfying $h^{ab} h_{bc}={\delta^a}_c$.

From Eqs.~\eqref{gamma}, \eqref{lagrangian} and~\eqref{EMtensor}, it follows that the trace of the $p$-brane energy-momentum tensor ($T\equiv T^{\alpha\beta}g_{\alpha\beta}$) is equal to 
\begin{eqnarray}
T_{\rm [\it p \rm]}&=&  -\mu_{\rm [\it p \rm]} \int  \sqrt{-h}  h^{ab} h_{ab}   \delta^{(N+1)} \left[x^\sigma - x^\sigma (\zeta)\right]d^{p+1} \zeta \nonumber \\
&=&  (p+1) \mathcal{L}_{\rm [\it p \rm]}\,,
\label{Relation_L_T}
\end{eqnarray}
where we have used the identity $h^{ab} h_{ab}=p+1$. Therefore, the Lagrangian of a Nambu-Goto $p$-brane is equal to $\mathcal{L}_{\rm [\it p \rm]}=T_{\rm [\it p \rm]}/(p+1)$, independently of the particular properties of the gravitational field. While for $p=0$ this reduces to the standard result $\mathcal{L}_{\rm [0]}=T_{\rm [0]}$, which defines the form of the on-shell Lagrangian of point particles and their associated fluids, more generally it depends on the microscopic properties of the matter source.

\section{Time-Averaged Lagrangian-Mass Relation for Cosmic String Loops\label{sec3}}

Here, we consider oscillating cosmic string loops, corresponding to closed 1-branes evolving in Minkowski spacetime (see \cite{Vilenkin:2000jqa} a detailed overview of cosmic string dynamics). Restricting our analysis to periodic loop configurations that do not self-intersect, we shall derive a relation between the spatial volume integral of their time-averaged Lagrangian and their rest mass. For simplicity, here and in the next section we suppress the subscript [1] from all quantities, except those involving the cosmic string Lagrangian. In particular, throughout these two sections we use the notation $\mu \equiv \mu_{\rm [1]}$, $T^{\alpha \beta} \equiv T^{\alpha \beta}_{\rm [1]}$, $T \equiv T_{\rm [1]}$.

Consider the worldsheet of a Nambu-Goto string, described by the embedding
\begin{equation}
x^\alpha = x^\alpha(\zeta) \,,
\end{equation}
where $\zeta^a$ ($a=0,1$) are the worldsheet coordinates, with $\zeta^0$ timelike and $\zeta^1$ spacelike.
In Minkowski space, one can adopt the conformal gauge, in which the induced metric of the string ($h_{ab}\equiv\gamma_{ab}$) satisfies
\begin{equation}
\gamma_{ab}  = \sqrt{-\gamma}  \eta_{ab} \,, \qquad \gamma^{ab}  = \frac{1}{\sqrt{-\gamma} } \eta^{ab} \,,
\label{gammaab} 
\end{equation}
where $\eta_{ab}$ is the ($1+1$)-dimensional Minkowski metric, and one may further choose $\zeta^0 = x^0 \equiv t$ as the timelike worldsheet coordinate. In this gauge, in three spatial dimensions the string position at time $t$ can be described by a three-dimensional vector field 
\begin{equation}
\vec{x}(\xi) = [x^1(\xi), x^2(\xi), x^3(\xi)]\,,
\end{equation}
which depends only on the spatial parameter $\xi \equiv \zeta^1$.  

The relation between the induced and background metrics, given in Eq.~\eqref{gammaab}, then implies the conformal gauge conditions on the string worldsheet
\begin{equation}
\dot {\vec x} \cdot \vec x \, '  =0\,, \qquad |{\dot {\vec x}}|^2 +  |{\vec x \, '}|^2  =1\,, \label{conformal} 
\end{equation}
where the dot and prime denote derivatives with respect to $t$ and $\xi$, respectively. The resulting equations of motion reduce to the standard wave equation on the string,
\begin{equation}
\ddot {\vec x} - \vec x\, '' =0\,.
\end{equation}
In the conformal gauge, the energy-momentum tensor of a Nambu-Goto string takes the form
\begin{equation}
T^{\alpha \beta} (t,\vec{x}) = \mu \int \left( \dot{x}^\alpha \dot{x}^\beta - x^{\alpha\prime} x^{\beta\prime} \right) \delta^{(3)} \left[\vec{x} - \vec{x}(t,\xi)\right] \, d\xi \,.
\end{equation}

A closed string loop (closed $1$-brane) of invariant length $L$ satisfies
\begin{equation}
\vec{x}(t,\xi + L) = \vec{x}(t,\xi) \,.
\end{equation}
Combined with the equations of motion, this implies that, in the center-of-mass frame, the motion is periodic in time with period
\begin{equation}
\Pi = \frac{L}{2} \,,
\end{equation}
and the rest mass of the loop is given by
\begin{eqnarray}
m
&=& - \int {T_{0}}^0 \, d^3 x \nonumber \\
&=& \mu \int_0^L
\left( \int \delta^{(3)} \!\left[\vec{x} - \vec{x}(t,\xi)\right]
\, d^3 x \right)
d\xi \nonumber \\
&=& \mu \int_0^L d\xi
= \mu L \,.
\label{mass}
\end{eqnarray}
In this frame, for each point on the loop, the velocity averaged over one period vanishes,
\begin{equation}
\langle \dot{\vec{x}} \rangle
= \frac{1}{\Pi} \int_0^\Pi \dot{\vec{x}}(t,\xi)\, dt
= \vec{0} \,.
\end{equation}
Consequently, in the center-of-mass frame, the time-averaged linear momentum of a non-self-intersecting cosmic string loop also vanishes,
\begin{equation}
\int \langle T^{0 i} \rangle \, d^3 x = \mu \int_0^L \langle \dot{x}^i \rangle \, d\xi = 0 \,.
\end{equation}
Moreover,
\begin{eqnarray}
\int \langle T^{i j} \rangle \, d^3 x
&=& \frac{1}{\Pi} \int_0^\Pi \int_0^L
\left( \dot{x}^i \dot{x}^j - x^{i\prime} x^{j\prime} \right)
\, d\xi \, dt \nonumber \\
&=& - \frac{1}{\Pi} \int_0^\Pi \int_0^L
x^i \left( \ddot{x}^j - x^{j\prime\prime} \right)
\, d\xi \, dt \nonumber \\
&=& 0 \,,
\label{int_T^ij}
\end{eqnarray}
where we have integrated by parts with respect to both $t$ and $\xi$ (the boundary terms vanish due to the periodicity of the loop in both space and time). Equation~\eqref{int_T^ij} implies that, in Minkowski spacetime, the spatial volume average of the time-averaged pressure of a cosmic string loop vanishes in the center-of-mass frame,
\begin{equation}
\mathfrak P\equiv \frac{\int \langle {T_{i}}^{i} \rangle d^3 x}{3}=0\,.
\label{pp}
\end{equation}

It then follows that
\begin{eqnarray}
\mathfrak T &\equiv& \int \langle T \rangle \, d^3 x
= \int \langle {T_{0}}^0 \rangle \, d^3 x
+ \int \langle {T_{i}}^i \rangle \, d^3 x \nonumber \\
&=& -m
+ 3 \, \mathfrak P = - m \,.
\end{eqnarray}
Using the relation between the 1-brane Lagrangian and the trace of the energy-momentum tensor derived in the previous section [Eq.~\eqref{Relation_L_T}], we further obtain
\begin{equation}
\mathfrak L \equiv \int \langle \mathcal{L}_{\rm [1]} \rangle \, d^3 x
= \frac12 \int \langle T \rangle \, d^3 x \equiv \frac{\mathfrak T}{2}
= - \frac{m}{2} \,.
\label{L=-m/2}
\end{equation}

\section{Cosmological Evolution of the Mass of Small Cosmic String Loops\label{cosmological}}

In this section, we show that the rest mass of a cosmic string loop in an FLRW background may evolve over cosmological timescales due to the nonminimal matter-geometry coupling. 
We will consider a regime where the string tension is sufficiently small that gravitational radiation emission by the cosmic string loop can be neglected. We further assume that the loop size is much smaller than the Hubble radius, so that the dynamics over a single oscillation are approximately unaffected relative to Minkowski spacetime.

Consider a flat homogeneous and isotropic universe described by the FLRW metric
\begin{equation}
    ds^2 = -dt^2 + d\vec{x}\cdot d\vec{x}\equiv -dt^2 + a^2(t) d\vec{q}\cdot d\vec{q} \, ,
\label{FLRW_metric}    
\end{equation}
where $t$ denotes cosmic time, $a(t)$ is the scale factor, and $\vec{x}$ and $\vec{q}$ represent physical and comoving Cartesian coordinates, respectively. In an FLRW spacetime, the $\alpha=0$ component of Eq.~\eqref{Continuity_Eq} becomes
\begin{equation}
\partial_0{T_{0}}^0+\partial_i{T_{0}}^i+3H{T_{0}}^0-H{T_{i}}^i=\frac{\dot \psi}{\psi}(\mathcal{L}_{\rm [1]}-{T_{0}}^0)\,,
\label{Continuity_Eq_FLRW}
\end{equation}
where $\psi=\psi(t)$, $H(t)\equiv \dot a/a$ is the Hubble parameter, and a dot denotes a derivative with respect to $t$.

Consider an isolated, non-self-intersecting cosmic string loop whose center of mass is at rest with respect to the cosmological frame. Its rest mass is given by the spatial volume integral of the energy density,
\begin{equation}
m=-\int {T_{0}}^0\, d^3x
=-a^3\int {T_{0}}^0\, d^3q \,.
\label{Def_m}
\end{equation}

Integrating Eq.~\eqref{Continuity_Eq_FLRW} over a spatial volume containing the loop, noting that the energy-momentum tensor vanishes outside it, so that
\begin{equation}
\int  \partial_i {T_\alpha}^\beta d^3 x =0\,,
\end{equation}
and using Eq.~\eqref{Def_m}, we obtain 
\begin{equation}
\dot{m}+ H \int {T_{i}}^i  d^3 x=- \frac{\dot{\psi}}{\psi} \int\left[ \mathcal{L}_{\rm [1]}  -{T_0}^0\right]d^3x \,.
\label{Mass_Conservation}
\end{equation}

Because the loop size $L$ is much smaller than the Hubble radius $H^{-1}$, the effect of cosmological expansion on the loop dynamics during a single oscillation is negligible. The motion of the loop is therefore quasi-periodic, with quasi-period
\begin{equation}
\Pi = \frac{L}{2}=\frac{m}{2\mu}\, .
\end{equation}
This separation between the short oscillation timescale and the much longer cosmological timescale  motivates describing the mass evolution in terms of quantities averaged over one quasi-period,
\begin{equation}
\dot{m}+H \int \langle {T^{i}}_i \rangle d^3 x=- \frac{\dot{\psi}}{\psi}\left[\int \langle \mathcal{L}_{\rm [1]} \rangle d^3  x+ m\right] \,,\label{Mass_Conservation_1}
\end{equation}
or, equivalently,
\begin{equation}
\dot{m}+3H \mathfrak P = - \frac{\dot{\psi}}{\psi}\left[\mathfrak L+ m\right] \,.
\label{Mass_Conservation_11}
\end{equation}
Here, 
\begin{equation}
\langle Q \rangle (t) \equiv \frac{1}{\Pi}\int_t^{t+\Pi} Q(\tau)\, d\tau
\end{equation}
denotes the average of the physical quantity $Q$ over a quasi-period $\Pi$. In deriving Eqs.~\eqref{Mass_Conservation_1} and~\eqref{Mass_Conservation_11} we neglected the tiny variations of $m$ occurring on timescales shorter than $\Pi$ (given that $\Pi \ll  |\psi/\dot \psi|$ and $\Pi \ll H^{-1}$, both the scale factor and $\psi$, are essentially constant over a time interval $\Pi$). Neglecting the effect of cosmological expansion on the loop dynamics during a single oscillation, Eqs.~\eqref{pp} and~\eqref{L=-m/2} imply that $\mathfrak P=0$ and $\mathfrak L=-m/2$. Substituting these into Eq.~\eqref{Mass_Conservation_11}, we finally obtain
\begin{equation}
\frac{\dot{m}}{m}=- \frac12 \frac{\dot{\psi}}{\psi} \,.
\label{Mass_Conservation_Final}
\end{equation}

Equation~\eqref{Mass_Conservation_Final} implies that the rest mass of the loop evolves as
\begin{equation}
m \propto |\psi|^{-1/2}\,.
\label{m(psi)}
\end{equation}
This result depends crucially on the relation $\mathfrak L_{\rm [1]}=-m/2$ derived for cosmic string loops. In contrast, for point particles with $\mathcal{L}_{\rm [0]} =T_{\rm [0]}$ one has $\mathfrak L_{\rm [0]}\equiv \int \langle \mathcal L_{\rm [0]} \rangle d^3 x=-m_{\rm [0]}$, which implies  $\dot m_{\rm [0]}=0$ in the analogue of Eq.~\eqref{Mass_Conservation_11} for point particles and thus conservation of the particle rest mass~\cite{Avelino_2022}. While in GR the result is the same for point particles and small cosmic string loops (the rest mass is conserved in both cases, despite their different physical nature), in nonminimally coupled scalar-tensor gravity, the rest mass of a cosmic string loop is directly coupled to the scalar field $\psi$. 

Since Eq.~\eqref{m(psi)} follows solely from Eq.~\eqref{Continuity_Eq}, it applies to all the subclasses introduced in Sec.~\ref{ST_theories}. In particular, for nonminimally coupled Brans-Dicke theory one has $\psi(t)=h[\phi(t)]$. Likewise, for
\begin{equation}
f(R,\mathcal{L}_{\rm m})=f_1(R)+f_2(R)\mathcal{L}_{\rm m},
\end{equation}
one has $\psi=f_2(R)>0$. In an FLRW universe, the Ricci scalar,
\begin{equation}
R=6(\dot H+2H^2),
\end{equation}
depends only on cosmic time, so that $\psi=\psi(t)$ and
\begin{equation}
m\propto f_2^{-1/2}\,.
\end{equation}
Hence, in both subclasses the rest mass of a cosmic string loop evolves in response to the cosmological evolution of the background, even when its size and string tension are arbitrarily small, whereas the rest mass of a point particle remains conserved.

\section{Cosmological Evolution of the Mass of Small Closed $p$-Branes\label{cosmological2}}

In this section, we study the cosmological evolution of the rest mass of small, closed $p$-branes whose center of mass is comoving with the cosmological expansion within the framework presented in Section \ref{ST_theories}. Building upon the general Lagrangian identity for Nambu-Goto $p$-branes, we investigate isolated, non-self-intersecting configurations in $(N+1)$-dimensional FLRW spacetimes. Again, we shall assume that the $p$-brane tension is sufficiently low for gravitational radiation emission to be negligible.

In \cite{Avelino:2008mv} it was shown that the average pressure of periodic, maximally symmetric Nambu-Goto $p$-branes in a de Sitter universe vanishes,
\begin{equation}
\mathfrak P_{\rm [\it p \rm]} \equiv \frac{\int \langle { T_{i}}^{i}_{\rm [\it p \rm]}\rangle d^N \! x}{p}=0\,,\label{mathpp}
\end{equation}
assuming the branes are isolated and do not experience self-interactions, either gravitational or of other types (here, $d^N\!x=a(t)^N d^N\!q$ is the $N$-dimensional physical volume element and $d^N\!q$ is the corresponding $N$-dimensional comoving volume element). Although this result strictly applies only to this idealized class of solutions, it is expected to provide an excellent approximation for sufficiently small, mass-conserving, non-self-intersecting $p$-branes in an FLRW spacetime in $(N+1)$-dimensional GR, as the energy density of a cosmological fluid composed of such $p$-branes branes scales as $\rho \propto a^{-N}$, consistently with a vanishing average pressure \cite{Avelino:2008mv}. Equation~\eqref{mathpp} is also a necessary condition for a closed $p$-brane to behave, on average, as a particle, in the sense that the standard Lorentz transformation laws for its energy and linear momentum are satisfied \cite{Giulini:2018tuw,Avelino:2022eqm}.

Intersections of $p$-branes are constrained by the transversality theorem in differential topology \cite{Guillemin:1974}, which states that the generic intersection of two submanifolds of dimensions $p_1$ and $p_2$ embedded in an $N$-dimensional space has dimension $p_1 + p_2 - N$. In particular, for two $p$-branes ($p_1 = p_2 = p$), the intersection generically has dimension $2p - N$, so that if $2p - N < 0$, intersections are not generic. This implies that for $p < N/2$, non-self-intersecting solutions with vanishing average pressure could, in principle, exist.

The results of the previous section can now be generalized to small, oscillating, non-self-intersecting $p$-branes background with rest mass
\begin{equation}
m_{\rm [\it p \rm]}=-\int {T_{0}}^0_{\rm [\it p \rm]} \, d^N\!x= - a^N \int {T_{0}}^0_{\rm [\it p \rm]}\,  d^N\!q \, ,
\label{Def_mp}
\end{equation}
in an $(N+1)$-dimensional FLRW background. Following an analogous procedure to that of the previous section, and assuming that the $p$-brane is sufficiently small so that the tiny variations of $m_{\rm [\it p \rm]}$ on timescales much shorter than $|\psi/\dot \psi|$ and $H^{-1}$ can be neglected, we find
\begin{equation}
\dot{m}_{\rm [\it p \rm]}+p H\, \mathfrak P_{\rm [\it p \rm]} = - \frac{\dot{\psi}}{\psi}\left[m_{\rm [\it p \rm]} +\mathfrak L_{\rm [\it p \rm]}\right] \,,
\label{Mass_Conservation_1p}
\end{equation}
with
\begin{eqnarray}
\mathfrak L_{\rm [\it p \rm]} &\equiv&  \int\langle \mathcal{L}_{\rm [\it p \rm]} \rangle \, d^N\!x = \frac{\mathfrak T_{\rm [\it p \rm]}}{p+1} \,,\\
\mathfrak T_{\rm [\it p \rm]} &\equiv& \int \langle T_{\rm [\it p \rm]} \rangle \, d^N\!x = -m_{\rm [\it p \rm]}+ p \, \mathfrak P_{\rm [\it p \rm]}\,.
\end{eqnarray}

Under these assumptions, Eq.~\eqref{mathpp} is expected to hold to an excellent approximation on timescales much shorter than both $|\psi/\dot \psi|$ and $H^{-1}$. In this regime,
\begin{equation}
\mathfrak L_{\rm [\it p \rm]}=-\frac{m_{\rm [\it p \rm]}}{p+1}\,,
\end{equation}
and the second term on the left-hand side of Eq.~\eqref{Mass_Conservation_1p} can be safely neglected. Hence, Eq.~\eqref{Mass_Conservation_1p} reduces to
\begin{equation}
\frac{\dot{m}_{\rm [\it p \rm]}}{m_{\rm [\it p \rm]}}=- \lambda\frac{\dot{\psi}}{\psi}\,, \qquad \lambda = \frac{p}{p+1} \, .
\label{Mass_Conservation_p-brane}
\end{equation}

Consequently, the rest mass of small, oscillating, low-tension, non-self-intersecting $p$-branes in an $(N+1)$-dimensional FLRW background evolves as
\begin{equation}
m_{\rm [\it p \rm]} \propto {|\psi|}^{-\lambda}\,. \label{mp}
\end{equation}
Equation~\eqref{mp} generalizes the results obtained in \cite{Avelino_2022} and in the previous section for $0$- and $1$-branes, respectively, with the exponent $\lambda$ varying between $0$ ($p=0$) and $1$ (in the $p\to\infty$ limit). For $p=1$, we recover the case of the cosmic string loop discussed in the previous section, for which $\lambda = 1/2$.

\section{Conclusions \label{concl}}

In this work, we investigated the evolution of the rest mass of particle-like objects in a broad class of scalar–tensor theories in which the matter Lagrangian is nonminimally coupled to a scalar field $\psi$, encompassing several well-known modified gravity theories as particular cases.

We showed that the rest mass of closed cosmic string loops, and more generally of closed $p$-branes with $p\ge1$, can evolve over cosmological timescales as a consequence of the nonminimal matter coupling, even when their size and tension are arbitrarily small. This behavior contrasts sharply with that of point particles, whose rest mass remains conserved within the same class of theories, as well as with small particle-like objects in general relativity, whose rest mass is insensitive to the cosmological evolution of the background spacetime.

The origin of this qualitative difference lies in the Lagrangian identity $\mathcal{L}_{[p]}=T_{[p]}/(p+1)$, which generalizes the familiar relation $\mathcal{L}_{\rm m}=T$ satisfied by point particles. We have shown that the conservation of the rest mass is intimately connected to this identity: in theories with nonminimal matter couplings, the rest mass remains constant only if the relation $\mathcal{L}_{\rm m}=T$ holds. Consequently, the evolution of the rest mass provides a direct manifestation of the internal structure and dimensionality of the localized object.

We emphasize that this effect can remain physically meaningful even if the nonminimal coupling is significant only during a finite stage of the cosmological evolution. Indeed, the coupling function may be tuned so that the nonminimal interaction is effectively absent at both early and late times, while remaining active during an intermediate epoch. In such a scenario, two branes with identical initial rest masses may emerge with different final rest masses once the interaction has ceased. Since the evolution law depends explicitly on the dimensionality $p$ of the brane, these final masses likewise depend on $p$, despite the identical initial conditions. Thus, the evolution of the rest mass is not merely a transient consequence of the nonminimal interaction but may leave a permanent imprint that persists long after the interaction has ceased.

These results have important implications for the modeling of particle-like objects in theories with nonminimal matter couplings. In particular, they show that the cosmological evolution of their mass depends not only on the underlying theory of gravity but also on their internal structure. This dependence may have significant consequences in string-inspired theories, where extended objects, such as strings or higher-dimensional branes, play a fundamental role. More broadly, our analysis highlights the link between the microphysics of particle-like objects and cosmological dynamics in theories with nonminimal matter-gravity couplings, with potential applications ranging from the dynamics of topological defects to the effective description of particle-like objects and the cosmological evolution of the Universe in theories beyond GR.

\acknowledgments

We thank our colleagues of the Cosmology group at Instituto de Astrofísica e Ciências do Espaço for enlightening discussions and acknowledge the support by Fundação para a Ciência e a Tecnologia (FCT) through the research grants No. UID/04434/2025 (DOI: \href{https://doi.org/10.54499/UID/04434/2025}{10.54499/UID/04434/2025}) and No. 2024.17828.PEX -- \emph{Unveiling the Early Universe with Topological Defects} (DOI: \href{https://doi.org/10.54499/2024.17828.PEX}{10.54499/2024.17828.PEX}). S. R. P. also acknowledges the support by Fundação para a Ciência e a Tecnologia (FCT) through the grant No. 2025.03891.BD.

\bibliography{article}

\end{document}